# Tuning hole mobility of individual p-doped GaAs nanowires by uniaxial tensile stress

Lunjie Zeng[1*], Jonatan Holmér[1], Rohan Dhall[2], Christoph Gammer[3], Andrew M. Minor[2,4], and Eva Olsson[1*]

1 Department of Physics, Chalmers University of Technology, 412 96 Gothenburg, Sweden

2 National Center for Electron Microscopy, Molecular Foundry, Lawrence Berkeley National Laboratory, Berkeley, California 94720, United States

3 Erich Schmid Institute of Materials Science, Austrian Academy of Sciences, 8700 Leoben, Austria

4 Department of Materials Science and Engineering, University of California, Berkeley, California 94720, United States

*Corresponding author:

lunjie@chalmers.se

eva.olsson@chalmers.se

**Abstract**

Strain engineering provides an effective way of tailoring the electronic and optoelectronic properties of semiconductor nanomaterials and nanodevices, giving rise to novel functionalities. Here, we present direct experimental evidence of strain-induced modifications of hole mobility in individual GaAs nanowires, using in situ transmission electron microscopy (TEM). The conductivity of the nanowires varied with applied uniaxial tensile stress, showing an initial decrease of ~5-20% up to a stress of 1~ 2 GPa, subsequently increasing up to the elastic limit of the nanowires. This is attributed to a hole mobility variation due to changes in the valence band structure caused by stress and strain. The corresponding lattice strain in the nanowires was quantified by in situ 4D-scanning TEM (STEM) and showed a complex spatial distribution at all stress levels. Meanwhile, a significant red shift of the band gap induced by the stress and strain was unveiled by monochromated electron energy loss spectroscopy.

III-V compound semiconductors possess unique optical and electronic properties, making them important components for advanced devices, such as transistors, lasers, sensors, and solar cells [1]. Among III-V semiconductors, GaAs exhibits outstanding charge transport and optical characteristics, like high charge carrier mobility and ability to efficiently detect and emit light [2]. These properties have been utilized to fabricate high-speed electronics, high-efficiency solar cells and sensitive detectors for optical communication [3]. For further enhancing the performance of these devices for future generation electronics and optoelectronics, two strategies are usually adopted. One approach is to downscale the physical size of the basic components so that more signal processing power can be integrated into the devices. This has initiated an enormous research interest in nanoscale semiconductor materials. GaAs nanostructures, especially nanowires, have been fabricated and demonstrated an attractive potential for performance enhancement [4–7]. The other strategy is to optimize material functionalities by engineering material structure through, for instance, mechanical strain. The coupling between mechanical stress/strain and electronic and optoelectronic properties of semiconductors has attracted extensive research interest since the discovery of piezoresistance effect in Si [8–11]. Strain engineering has also been shown to have a distinct effect on the properties of bulk GaAs [12–14]. The understanding of strain effect on charge transport in GaAs nanowires is thus of great importance for developing nanoscale semiconductor devices with improved and novel functionalities.

Distinct strain effects on electronic structure and electrical as well as optical properties of GaAs and related semiconductor nanomaterials have been found previously. Pressure induced band structure modification has been observed in InP nanowires [15]. Charge transport properties of individual InAsP and InAs nanowires have shown a sensitive response to mechanical stress [16,17]. Light emission in Zinc Blende (ZB) GaAs nanowires has been tuned by uniaxial compressive and tensile strain over a large spectrum range [18]. A large variation in the band gap of GaAs nanowires due to lattice-mismatch strain has been observed in GaAs/In$_x$Ga$_{1-x}$As core/shell nanowires [19]. A direct to indirect band gap transition introduced by uniaxial stress and strain has been reported in Wurtzite GaAs nanowires [20]. The effect of bending deformation on charge transport in GaAs nanowires has also been investigated [21]. Hole transport characteristics in GaAs are critical for its application in field effect transistors and solar cells [4,22], but the strain effect on hole transport in GaAs nanowires is not fully understood due to the complexity in their valence band structure [10]. Controversial results have been theoretically predicted [10,23–26]. In this work, we studied the effect of uniaxial tensile stress on transport properties in individual p-doped GaAs nanowires. With an in situ transmission electron microscopy (TEM) setup, we applied tensile stress on the nanowires along the length direction up to ~5 GPa. The resulting lattice strain in the nanowires was quantified using scanning TEM – nanobeam electron diffraction (STEM-NBED, or 4D STEM) strain mapping at the nanometer resolution. The I-V characteristics of the nanowires were also simultaneously measured. The changes in the I-V characteristics were attributed to the strain induced changes in hole mobility that originated from electronic band structure modification. The band structure changes were verified by tight-binding simulations. In addition, a red shift of band gap energy due to tensile strain was observed by in situ electron energy loss spectroscopy (EELS). The present investigation verifies the scenario of a competition between effective mass change and charge scattering rate modification due to strain in the GaAs nanowires. The findings may enable delicate tailoring of the performance of GaAs based electronic and optoelectronic nanodevices by mechanical stress.

The GaAs nanowires used in this study were grown on Si(111) substrate using a molecular beam epitaxy (MBE) system by a self-catalysed vapor-liquid-solid (VLS) method [27]. The in situ tensile test and electrical transport properties measurements on the nanowires in TEM were enabled by a Hysitron PI95 nanoindenter TEM holder with an electrical push-to-pull (EPTP) microelectromechanical system (MEMS) device (Supporting Information S1) [16,17,28].

Several nanowires were investigated with consistent results. Details of results from one nanowire (Nanowire 1) are described below. Additional data from other nanowires are provided in Supporting Information. The mechanical stress applied on the nanowire along the length direction through the push-to-pull mechanism is extracted from the force – displacement characteristics of the nanoindenter in the in situ TEM holder. The indentation force – indenter displacement relationship was measured for the EPTP MEMS device without the nanowire and for the case where the nanowire was mounted on the MEMS device (Figure 1(a)). The indentation force increases linearly with indenter displacement for both cases. The linearity demonstrates that both the MEMS device and the nanowire deform elastically under the influence of the applied uniaxial tensile stress. When the total indentation force exceeded ~ 250 µN, the nanowire fractured. The tensile force applied on the nanowire was obtained by taking the difference between the forces applied on the MEMS device with and without the nanowire mounted on it. The diameter of the nanowire is around 160 nm according to STEM imaging. The tensile stress (**σ**) applied on the nanowire before the fracture was thus calculated and plotted as a function of indenter displacement (Figure 1(b)). At the fracture point, more than 5 GPa tensile force is applied on the nanowire. It is worth noting that such a stress level is much higher than that realized in GaAs bulk materials and in thin films [10,25,29].

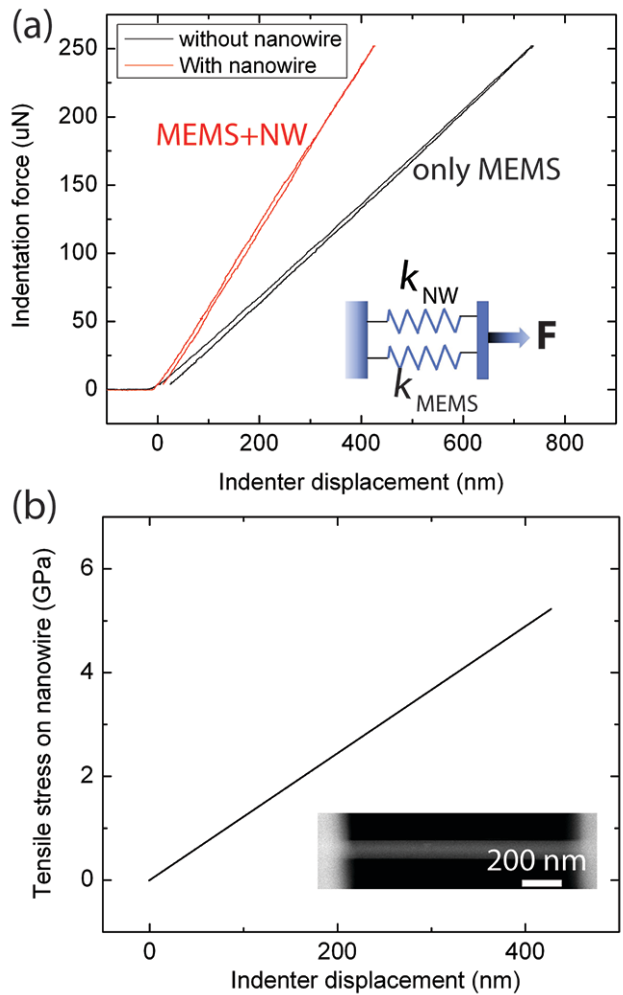

**Figure 1. Tensile stress applied on a GaAs nanowire through the push-to-pull mechanism.** (a) Indentation force applied on the MEMS device with (red curve) and without (black curve) GaAs nanowire on the device as a function of indentation distance. The inset is a schematic of two parallelly connected springs, which are analogue to the mechanical connection between the MEMS device and the nanowire, under tensile stress. $K_{NW}$ and $K_{MEMS}$ are the spring constants of the nanowire and EPTP MEMS device, respectively. F is the tensile force applied on the parallel springs. (b) Tensile stress ($\sigma$) applied on the nanowire plotted as a function of the indenter displacement. To calculate mechanical stress applied on the nanowire, tensile force applied on the nanowire was extracted from (a) and the diameter of the nanowire was measured from the STEM images. A representative STEM image of the nanowire on an EPTP MEMS device is shown as inset.

Lattice strain and its distribution in the GaAs nanowire are quantified by in situ STEM-NBED strain mapping. STEM-NBED measurements provide quantitative strain distribution information with high precision, high spatial resolution, and large field of view [30,31]. The nanowire has ZB crystal structure, according to electron diffraction. The [2-1-1] zone axis of the nanowire is aligned with the incident electron beam direction for STEM-NBED data collection (Supporting Information S2). Local strain within the nanowire is measured along the nanowire length direction ([111] direction or x direction) and the perpendicular direction ([02-2] direction or y direction). The resultant nanoscale 2D strain maps for $\varepsilon_{xx}$ and $\varepsilon_{yy}$ show the dynamic evolution of lattice strain over the nanowire due to applied uniaxial tensile stress. When there is no stress applied, most of the nanowire area is without lattice strain. As the applied uniaxial tensile stress increases, the nanowire is elongated. The general trend in the $\varepsilon_{xx}$ map is that more and more nanowire areas start to show tensile strain as the nanowire is stressed gradually, and there is always spatial inhomogeneity in the strain distribution within the nanowire. When the applied stress is around 1 GPa, most of the nanowire has a tensile strain along [111] direction, but there are still regions with little strain. The mean strain value $<\varepsilon_{xx}>$ and standard deviation $\sigma_{xx}$ are around 1.1% and 0.29%, respectively. We note that the uncertainty of the strain measurement is ~ 0.06% (Supporting Information S3). When the applied tensile stress reaches ~ 2.05 GPa, the whole nanowire is strained along the [111] direction. As the stress increases further, the $\varepsilon_{xx}$ map maintains a similar spatial strain distribution pattern until fracture. In the $\varepsilon_{yy}$ map, the compressive strain starts to emerge on the right side of the strain map when the stress is increased to ~ 1 GPa. At around 1 GPa, the mean strain value $<\varepsilon_{yy}>$ and standard deviation $\sigma_{yy}$ are around -0.2% and 0.29%, respectively. Then, the region with compressive strain increases in size and expands to the left side of the nanowire as the stress increases. Despite the local inhomogeneous strain distribution in the nanowire along both x and y directions, the absolute values of average strains $<\varepsilon_{xx}>$ and $<\varepsilon_{yy}>$ increase linearly with applied stress until fracture, showing elastic deformation of the nanowire structure under tensile stress (Figure 2(c)). This is consistent with the linear indentation force – indenter displacement characteristics shown in Figure 1. By fitting the tensile strain – tensile stress ($<\varepsilon_{xx}>$ – **σ**) data in Figure 2 (c) to a linear function, Young's modulus of the nanowire is estimated to be ~ 105 GPa. The Poisson's ratio of the nanowire is taken as the ratio between $<\varepsilon_{yy}>$ and $<\varepsilon_{xx}>$, which is calculated to be ~ 0.22 (± 0.03). Young's modules of several GaAs nanowires have been measured and vary from ~ 90 GPa to ~120 GPa. In comparison, the bulk Young's modulus of GaAs along [111] direction is ~ 142 GPa [32]. The Poisson's ratio of bulk GaAs is around ~ 0.19 [32].

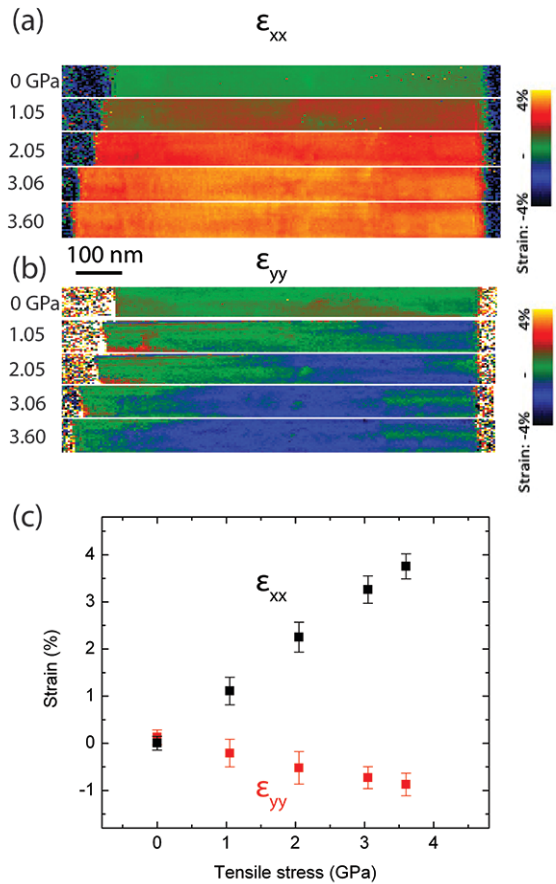

**Figure 2. In situ strain mapping by STEM-NBED.**

(a) Strain maps of the GaAs nanowire under tensile stress, with strain measured along the nanowire length (x) direction. From top to bottom, the tensile stresses applied on the nanowire are 0, 1.05, 2.05, 3.06, and 3.60 GPa, respectively.

(b) Strain maps of the GaAs nanowire under tensile stress, with strain measured perpendicular to the nanowire length (y) direction. From top to bottom, the tensile stresses applied on the nanowire are 0, 1.05, 2.05, 3.06, and 3.60 GPa, respectively.

(c) Strain distribution within the nanowire as a function of applied tensile stress. Data points are the mean strain values in the strain maps, and error bars correspond to the standard deviation of strain values in the strain maps. The Poisson's ratio of the nanowire is ~ 0.22 along the [111] (~ 0.19 for bulk GaAs). The Young's modulus of the nanowire is ~ 105 GPa ( ~ 142 GPa for bulk GaAs).

Mechanical stress and strain induce modifications in the charge transport properties of the nanowire. In the electrical measurement setup, the GaAs nanowire is connected to metal electrodes on the EPTP device, forming a metal-semiconductor-metal (M-S-M) structure. The I-V characteristics of the M-S-M structure are shown in Figure 3(a). I-V curves measured in the range from -4 V to 4 V show symmetrical and nonlinear characteristics. Such I-V characteristics are consistent with the model where the semiconductor nanowire is treated as a resistor and is sandwiched between two identical and head-to-head Schottky barriers [33,34]. In the relatively low bias and nonlinear region, the I-V behavior of the circuit is dominated by the reverse-biased Schottky barrier in the M-S-M structure. At high bias and high current region, the electrical current varies linearly with applied bias. In this region, the I-V relation is determined by the conductance of the semiconductor nanowire. The theoretical model used to describe the I-V relationship in the nanowire M-S-M structure is given in Supporting Information (Figure S4). Parameters, including the nanowire resistance (conductance) and the Schottky barrier heights, were extracted by fitting the model to the experimental I-V curves. Through such a quantitative analysis, the nanowire conductance can be reliably obtained by decoupling the effect of electric contacts on the I-V characteristics. When the applied tensile stress increases from 0 to ~ 1GPa, the conductance of the nanowire decreases about 5 %, from ~ 0.71 to ~ 0.67 1/MΩ. As the stress increases to ~ 2.05 GPa, the conductance increases to a value similar as that at 0 GPa. When the applied stress elevates further, the conductance of the nanowire gradually increases. The conductance shows an almost linear increase with stress after 2 GPa.

At ~ 3 GPa, the nanowire conductance becomes ~ 0.80 1/MΩ. Finally, the nanowire conductance reaches around 0.87 1/MΩ at ~ 4.21 GPa. The variation in conductance of the nanowire is reproducible based on the repeated measurements performed on the same nanowire. The conductance change due to dimension variation of the nanowire is negligible, so the predominant source of conductance change is the alternation in conductivity of the nanowire. The initial reduction of nanowire conductivity followed by an enlargement in conductivity is observed in all the p-GaAs nanowires we studied (e.g. Supporting Information S5 and S6). The initial dip in conductivity is as much as 20% of the original conductivity in some nanowires (Supporting Information Figure S5 and S6). The conductivity of the nanowires at their elastic limit varies from ~ 80% to ~ 120% of the original conductivity values.

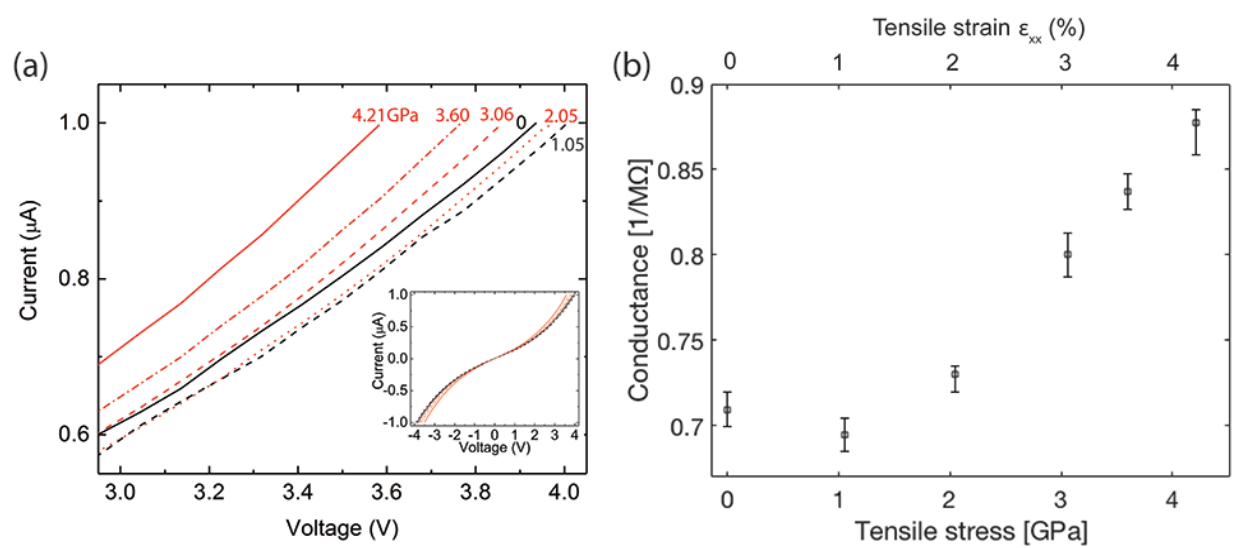

**Figure 3. Effect of mechanical stress and strain on the conductance of p-GaAs nanowire.**

(a) I-V characteristics of the GaAs nanowire under tensile stress. I-V curves are shown in the voltage range between 3.0 V and 4.0 V, where the linear region of the I-V curves is clearly visible. The tensile stresses applied on the nanowire corresponding to the I-V curves are labelled next to each I-V curve. Curves in black show a decrease in current and the slope of the linear part with applied tensile stress. Slope of the red curves gradually increases as the applied tensile force increases. The inset shows the I-V curves in the voltage range from -4.0 to 4.0 V.

(b) The conductance of the nanowire as a function of applied tensile stress and strain. Error bars correspond to uncertainties in the data fitting. Strain values are mean strain values in the strain maps.

To understand the changes in the conductivity of p-GaAs nanowires introduced by strain, its effect on the band structure is studied. Mechanical strain can cause shift, splitting and warping in energy bands due to changes in lattice symmetry, bond length and angles. A tight-binding simulation is conducted to model the effect of tensile strain along the [111] direction on the band edges at the conduction band bottom and valence band top in GaAs (Figure 4) [35,36]. Without strain, at the $\Gamma$ point, the bottom of the conduction band is formed by a single and almost parabolic band. The top of the valence band consists of a heavy hole (HH) band, a light hole (LH) band and a split-off band. HH and LH bands are degenerate at the $\Gamma$ point in the absence of strain. Under tensile strain, the conduction band shifts downwards in energy. The changes in valence band are more complicated. The degeneracy between the HH and LH bands is lifted due to strain, causing band splitting. The HH band moves upwards along energy axis, while the LH band shifts downwards. The split-off band also shifts downwards slightly. The shifts of the conduction, HH and LH bands vary almost linearly with tensile strain. These modifications in band structure of GaAs by tensile strain have also been predicted previously using the k.p method [10,18], and observed experimentally [37]. The splitting and shift of the band edges will result in a reduction in the band gap of GaAs, which is experimentally observed here by the in situ monochromated EELS measurements (Figure 4(b) and Figure S7). Under tensile stress and strain, the band gap onset in the EEL spectra shows a red shift. At zero strain, the band gap onset is around 1.4 eV, which is in line with the band gap values reported previously for ZB GaAs at room temperature [2]. As the average tensile strain within the nanowire grows, this band gap onset gradually shifts to lower energy. With around 3% tensile strain, the band gap energy decreases about 0.1 eV. The rate of band gap energy change as a function of applied stress is ~30 meV/GPa. Such a change in band gap as a function of strain and stress is consistent with that predicated by the band structure modelling (Figure 4(a)). In degenerate semiconductors, such as the p-doped GaAs nanowires used in this study, the carrier concentration is hardly affected by such a change in band gap. Thus, despite the band gap variation, the change in hole concentration in the nanowires due to strain is expected to be insignificant. As a result, the strain induced alternation in the conductivity of the p-GaAs nanowires should mainly originate from the modification of hole mobility.

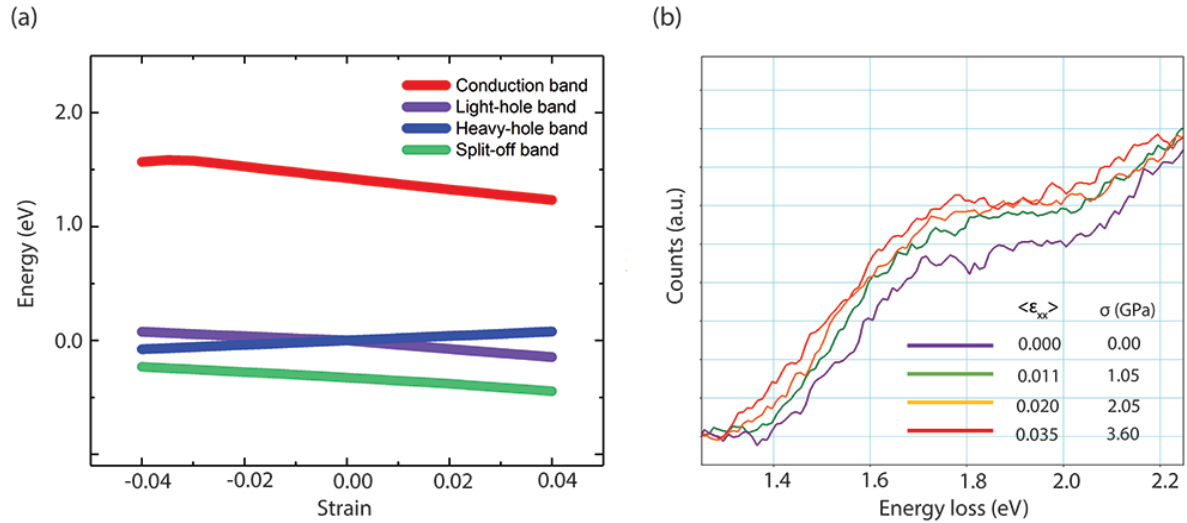

**Figure 4. Effect of uniaxial stress and strain on the band structure of GaAs nanowires**:

(a) Tight-binding simulation of band edges at the valence band top and conduction band bottom in GaAs and their shift as a function of strain along [111] direction.

(b) In situ monochromated EELS spectra showing the red shift of the band gap onset of the GaAs nanowire under stress and strain.

Tensile strain induced alternations in hole mobility in p-GaAs can be understood via the changes in valence band structure by taking into consideration the variations in hole effective mass and scattering rate. When there is no strain, the valence band top is a mixture of both the HH and LH bands (Figure 5(a)). The holes that contribute to charge transport reside on the top of the valence band and thus have the characteristics of both the HH and LH bands. Without strain, the hole effective masses are $m_{hh}^* = 0.5 m_0$ and $m_{lh}^* = 0.076 m_0$ [2,10]. The hole mobility is also determined by a variety of charge scattering processes in the material, among which polar optical phonon scattering is the dominant scattering mechanism at room temperature in GaAs. The optical phonon energy of GaAs, $h\omega_o$, is about $34\ meV$ [38]. Scattering of holes can happen within the same band as well as between HH and LH bands, facilitated by optical phonons. As described above, tensile strain causes splitting of the HH and LH bands, with the HH band moving up and LH band shifting down in energy. One consequence of such band shifts is that more free holes will reside in the HH band than in the LH band. Therefore, the average hole effective mass increases due to tensile strain. At relatively low strain levels, the band splitting energy is smaller than the optical phonon energy (Figure 5(b)). The joint density of states for interband scattering is almost the same as those in the unstrained case and thus phonon scattering of charge carriers changes insignificantly. Therefore, at small strain levels, the band shift and splitting gives rise to an increase in the average hole effective mass, resulting in a drop in the hole mobility and the nanowire conductivity. At high strain levels, more holes stay in the HH band and the hole effective mass keeps increasing. However, when the splitting between HH and LH bands is larger than the optical phonon energy, the phonon scattering of holes are largely suppressed due to the decline in joint density of states (Figure 5(c)). The reduced charge scattering increases hole mobility. If the reduction in phonon scattering outweighs the increase in hole effective mass, the hole mobility and hence the conductivity of the p-GaAs nanowire increase. Therefore, the reduction in hole mobility of the nanowire is predominantly determined by the increase in hole effective mass at low strain levels, while the enhancement of hole mobility is determined by both effective mass change and suppression in phonon scattering at high strain levels. Such a model is consistent with the theoretical work on metal-oxide-semiconductor field effect transistor (MOSFET) devices based on Si, Ge, as well as III-V semiconductors [10,24,26]. Despite the optical phonon energy in GaAs nanowires increases slightly (~ a few meV) with strain [18], the splitting between the heavy and light hole bands exceeds the optical phonon energy when the induced tensile strain is above about 0.8% according to our simulation (Figure 4(a)). This is consistent with our measurements, in which the minimum conductivity of the p-GaAs nanowires is reached with a uniaxial tensile strain of around 1~2%.

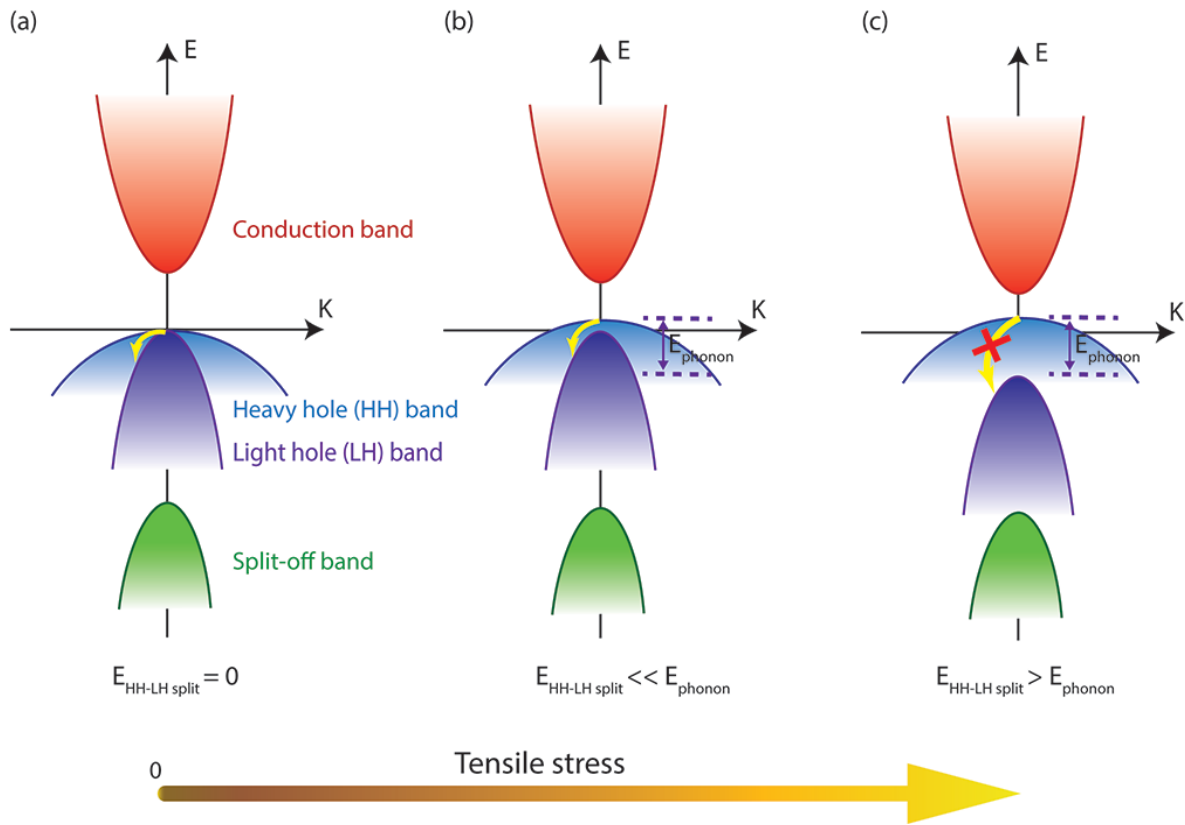

**Figure 5. Band diagram illustrating strain induced hole mobility change in GaAs due to modification in valence band structure.**

(a) When the nanowire is not stressed, the top of the valence band is formed by heavy hole band and light hole band, which are degenerate at $\Gamma$ point. There is a spin-orbital split-off band below the HH and LH bands. The yellow arrow indicates interband scattering of holes from HH to LH band facilitated by optical phonons. When applying tensile stress, HH shifts up in energy while LH shifts downwards, resulting a split between the two bands, as shown in (b) and (c).

(b) At low stress levels, the HH-LH band splitting energy is small in comparison with the optical phonon energy of GaAs and the interband phonon scattering is therefore not much affected. Another consequence of the splitting of the HH and LH bands is that most of the holes reside in HH band. As a result, the effective mass of the charge carriers increases compared to the unstressed case, decreasing the conductivity of GaAs.

(c) At high stress levels, when the split between HH and LH bands is larger than optical phonon energy, the interband phonon scattering of charge carriers is largely suppressed. Such a reduction in carrier scattering dominates the effective mass change, resulting an increase in conductivity.

In summary, we have directly and quantitatively investigated the uniaxial tensile stress and strain induced modification of conductivity in individual p-GaAs nanowires by using in situ TEM. Tensile stresses up to ~ 5 GPa are applied on individual nanowires through a push-to-pull mechanism in combination with nanoindentation. The induced strain in the nanowire has been quantified using 4D STEM strain mapping. Despite spatial strain inhomogeneity within the nanowire, the average lattice strain varies linearly with stress before nanowire fractures, with a Young's modulus value of around 105 GPa. The conductivity of the nanowires first reduces about 5-20% under tensile strains of ~ 1-2% (tensile stress ~ 1-2 GPa), whereafter it gradually increases as the stress and strain increases. Such an unusual change in nanowire conductivity is due to the alternation in the hole mobility of the nanowires, which originates from strain-induced modification of the valence band structure. This study helps improving our understanding of the intriguing correlation between lattice deformation, band structure variation and charge transport in semiconductor nanomaterials. It also demonstrates that mechanical strain can be used to delicately tailor electronic and optoelectronic properties of nanoscale materials.

**Acknowledgements**

L.J.Z., J.H. and E.O. acknowledge the financial support from Swedish Research Council (VR) under grant no. 2016-04618 and Excellence Initiative Nano (EI Nano) at Chalmers University of Technology. This project has also received funding from the European Union's Horizon 2020 research and innovation program under grant agreement No.823717-ESTEEM3. The authors acknowledge Professor Peter Krogstrup for providing the nanowire samples. Work at the Molecular Foundry was supported by the Office of Science, Office of Basic Energy Sciences, of the U.S. Department of Energy under Contract No. DE-AC02-05CH11231.

**Supporting Information**

Experiments and methods, NBED pattern, uncertainty in the STEM-NBED strain measurement, extracting electrical transport property parameters in a nanoscale metal-semiconductor-metal system by quantitatively analyzing I-V characteristics, repeated straining and I-V measurements on Nanowire 2, in situ straining and I-V measurements on Nanowire 3 using a scanning tunneling microscope – scanning electron microscope (STM-SEM) setup, in situ monochromated valence EELS of the GaAs nanowire under tensile stress, effect of tensile stress on bulk plasmon excitation in GaAs nanowire.